\documentclass[aps,final,notitlepage,oneside,twocolumn,nobibnotes,nofootinbib,
superscriptaddress,noshowpacs,centertags]{revtex4-1}

\usepackage[cp1251]{inputenc}
\usepackage[english]{babel}
\usepackage{graphicx}
\usepackage{latexsym}
\usepackage{amssymb}
\usepackage{amsmath}
\usepackage{float}

\begin{document}

\title{Dark Matter Density Spikes around Primordial Black Holes}
\author{Yu. N. Eroshenko}\thanks{e-mail: eroshenko@inr.ac.ru}
\affiliation{Institute for Nuclear Research, Russian Academy of Sciences,
pr. 60-letiya Oktyabrya 7a, Moscow, 117312 Russia}

\date{\today}

\begin{abstract}
We show that density spikes begin to form from dark matter particles around primordial black
holes immediately after their formation at the radiation-dominated cosmological stage. This follows from
the fact that in the thermal velocity distribution of particles there are particles with low velocities that
remain in finite orbits around black holes and are not involved in the cosmological expansion. The
accumulation of such particles near black holes gives rise to density spikes. These spikes are considerably
denser than those that are formed later by the mechanism of secondary accretion. The density spikes must
be bright gamma-ray sources. Comparison of the calculated signal from particle annihilation with the
Fermi-LAT data constrains the present-day cosmological density parameter for primordial black holes with
masses $M_{\rm BH}\geq10^{-8}M_\odot$ from above by values from $\Omega_{\rm BH}\leq1$ to $\Omega_{\rm BH}\leq10^{-8}$, depending on $M_{\rm BH}$. These
constraints are several orders of magnitude more stringent than other known constraints.
\end{abstract}

\maketitle 




\section{Introduction}

Primordial black holes (PBHs), the possibility
of whose formation was predicted in \cite{p41} and \cite{p22}, can give
valuable information about processes in the early
Universe \cite{p12,p14,p3}, in particular, about the shape of the perturbation
spectrum on small scales \cite{p24}.
The quantum evaporation of low-mass PBHs is
important from the viewpoint of investigating fundamental
processes at high energies \cite{p9} and can have significance for the
theory of primordial nucleosynthesis and gamma-ray
astronomy. In addition, PBHs can offer new
possibilities for the formation of quasars at high $z$
\cite{p16} and for baryonic objects with
chemical peculiarities\cite{p18,p19}. Being captured by neutron stars,
PBHs can affect their evolution, which gives a constraint
on the number of PBHs \cite{p10}.
In this paper, we will discuss only the PBHs that
are formed during the collapses of adiabatic density
perturbations, when a mixture of relativistic particles
collapses into a PBH at the instant the perturbation
crosses the cosmological horizon \cite{p11}. Note,
however, that other PBH formation models have also
been proposed at early dust-like stages \cite{p25,p40} or through
the collapses of domain walls \cite{p4},
\cite{p26,p37}.

PBHs can themselves represent dark matter (DM)
\cite{p23} if they are formed in sufficiently
large quantities, but they can also serve as seeds
for the formation of DM clumps \cite{p15,p35,p30,p36,p28,p38}. Secondary accretion (generally,
this mechanism was developed in cold DM
onto a PBH \cite{p7}, when DM flows
toward the PBH and is virialized at some radius to
form a halo, is usually considered in investigating DM
clumps around PBHs. In this paper, we will show
that the DM density around PBHs can reach much
greater values than that under secondary accretion.
This stems from the fact that in the thermal velocity
distribution there are DM particles with low velocities
that remain in finite orbits around PBHs and are not
involved in the overall cosmological expansion. The
accumulation of such particles around PBHs gives
rise to density spikes (halos).

Two regimes of density spike formation around
PBHs are possible at the radiation-dominated stage.
In the first case, which occurs for PBHs with masses
$M_{\rm BH}\leq40M_\odot$, PBHs are formed before the kinetic
decoupling of DM particles (under the assumption
that the DM particles are neutralinos with masses
$m\sim70$~GeV). In the interval between the PBH formation
and kinetic decoupling, a DM overdensity has
time to be formed around the PBH. As will be shown
below, the exact form of this initial density distribution
does not play a big role, while the separation of DM
particles immediately after their kinetic decoupling is
important. After their kinetic decoupling, the DM
particles begin to fly apart in the PBH gravitational
field, having some velocity distribution (a deformed
Maxwell distribution). Some of the particles with low
velocities remain gravitationally bound to the PBH,
forming subsequently a density spike around it. In
the second case, if $M_{\rm BH}>40M_\odot$, such a PBH is
formed already after the kinetic decoupling of DM
particles, and there is no initial overdensity of radiation
and DM around the PBH. In this case, the
DM particles with low velocities also remain in finite
orbits around the PBH, producing a density spike.
Thus, DM density spikes are formed around PBHs
at the radiation-dominated stage. After the onset
of the matter-dominated stage in the Universe, the
DM mass around PBHs begins to grow during the
secondary accretion, and a universal density profile
$\rho\propto r^{-9/4}$ is formed.

The DM density in the central regions of the spikes
is so large that by now the DM particles have managed
to annihilate (under the assumption that standard
neutralinos constitute the DM) at distances that
exceed the gravitational PBH radii by several orders
of magnitude. For this reason, to calculate the
present-day density profile around PBHs, it will be
sufficient for us to consider the phenomena at great
distances from the PBHs, where Newtonian gravitational
dynamics is a good approximation and the
general relativity effects are unimportant. The DM
remaining at great distances continues to annihilate
at present, producing signals in gamma-ray emission.
Comparison of the calculated signals with the
Fermi-LAT data allows the number of PBHs to be
constrained.

The annihilation of DM particles in clumps around
PBHs has already been considered in \cite{p28,p38,p17}, where constraints on the cosmological PBH
density parameter were obtained. Calculations \cite{p28} and \cite{p38} assumed the density
profile in the central region of a clump to be close
to $\rho\propto r^{-3/2}$, while \cite{p17} considered power-law
profiles $\rho\propto r^{-\alpha}$ with $\alpha=1.5-3$. The annihilation of DM in density cusps around black holes was considered in 
\cite{Sanetal10}, \cite{Sanetal11}, \cite{Sanetal12}, and new gamma-ray constraints were obtained. The goal of this
paper is to calculate the density profile in the central
region of DM clumps around PBHs by taking into
account the initial thermal velocity distribution of DM
particles after their kinetic decoupling. We will show
that the density profile has a more complex form than
$\rho\propto r^{-\alpha}$. Knowledge of the density profile allows one
to calculate the signals from DM annihilation around
PBHs more reliably and to obtain constraints on the
number of PBHs in the Universe. 


\section{Evolution of the density around PBHs before kinetic decoupling}
\label{evolsec}

Consider the PBH formation at the radiation-dominated
cosmological stage \cite{p11}, when the
equation of state for the matter in the Universe is
$p=\rho c^2/3$. A thermalized mixture of photons and
ultrarelativistic particles called radiation for short
collapses into a PBH. If nonrelativistic DM particles
are already present at this time in the Universe, then
they move in the overall gravitational potential and, in
addition, can interact with radiation. As an example,
consider DM particles in the form of neutralinos
with masses $m\simeq70$~GeV. At early times, when
the temperature was high, $T\geq0.05mc^2$, neutralinos
were in chemical equilibrium with radiation, i.e., the
production of neutralinos and their pair annihilation
were equiprobable. As the Universe cooled down,
neutralinos dropped out of chemical equilibrium with
radiation but still continued to efficiently interact with
it through scatterings. The neutralino gas temperature
was maintained at the radiation temperature
level, and neutralinos could be entrained by radiation
flows, for example, by the flow toward an accreting
PBH. Finally, on further cooling of the radiation
to some temperature $T_d$, whose value depends on
the character of elementary interactions, the kinetic
decoupling of DM particles from the radiation occurs
at the time $t_d$, and the DM particles subsequently
move freely only under the influence of gravitational
forces. We will find the fraction of the DM particles
that remain gravitationally bound to the PBH (have
finite orbits) as they fly apart in the next section, while
first it is necessary to discuss the initial DM density
profile around the PBH before kinetic decoupling.

The mechanism for the formation of a DM density
spike depends on the relation between the formation
time of the PBH determined by its mass and the time
$t_d$ dependent on the character of interaction between
DM particles and radiation. The PBH in some perturbed
region is formed at the instant $t_{\rm H}$ this region
crosses the cosmological horizon, which depends on
the total mass $M_{\rm H}$ of the matter inside this region:
\begin{equation}
t_{\rm H}\simeq\frac{GM_{\rm H}}{c^3}=2.6\times10^{-13}\left(\frac{M_{\rm BH}}{10^{-8}M_{\odot}}\right)\mbox{~s}. 
\label{th}
\end{equation}
We take into account the fact that the mass $M_{\rm BH}$
of the forming PBH in the model of \cite{p11} is
$M_{\rm BH}=M_{\rm H}/3^{3/2}$. The age of the Universe is related
to the radiation temperature as
\begin{equation}
t=\frac{2.4}{\sqrt{g_*}}\left(\frac{T}{1\mbox{~MeV}}\right)^{-2}\mbox{~s}, \label{ttime}
\end{equation}
where $g_*$ is the number of degrees of freedom; therefore,
the dependence of $M_{\rm BH}$ on $T$ at $t_{\rm H}$ is
\begin{equation}
M_{\rm BH}\simeq40\left(\frac{g_*}{10}\right)^{-1/2}\left(\frac{T}{27\mbox{~MeV}}\right)^{-2}M_\odot.
\label{mbhtbig}
\end{equation}
The normalization factor in (\ref{mbhtbig}) is chosen to correspond
to the temperature of the kinetic decoupling
of neutralinos with masses $m\simeq70$~GeV \cite{p6}
\begin{equation}
T_d\simeq27\left(\frac{m}{70\mbox{~GeV}}\right)^{1/4}\left(\frac{\tilde M}{0.2\mbox{~TeV}}\right)
\left(\frac{g_*}{10}\right)^{1/8}\mbox{MeV}, \label{tbigd}
\end{equation}
which occurs at a time
\begin{equation}
t_d\simeq10^{-3}\left(\frac{m}{70\mbox{~GeV}}\right)^{-1/2}\left(\frac{\tilde M}{0.2\mbox{~TeV}}\right)^{-2}\left(\frac{g_*}{10}\right)^{-3/4}\mbox{s}, \label{tsmd}
\end{equation}
where $\tilde{M}$ is the supersymmetry parameter \cite{p6}. Thus, the mass of $\sim40M_\odot$ given by
Eq.~(\ref{mbhtbig}) is a boundary value. If $M_{\rm BH}<40M_\odot$, then
the kinetic decoupling of neutralinos occurs already
after the PBH formation, while the radiation flow
accreted onto the PBH entrained DM particles from
the PBH formation time to $t_d$. If $M_{\rm BH}>40M_\odot$,
then the neutralinos at the PBH formation time were
free and moved independently from the radiation. The
radiation could outflow from some region of space,
while the DM remained in this region.

Consider the case of $M_{\rm BH}<40M_\odot$. We can
single out the near zone bounded by the radius of
influence of the PBH $r_{\rm infl}(t)$ in which the PBH mass is
equal to the radiation mass$M_{\rm BH}=(4\pi/3)\rho_{\infty}(t) r_{\rm infl}^3$,
where $\rho_{\infty}(t)=3/(32\pi Gt^2)$. Hence
\begin{equation}
r_{\rm infl}(t)=(8GM_{\rm BH}t^2)^{1/3}.
\label{inflrad1}
\end{equation}
In dimensionless units,
\begin{eqnarray}
\xi&=&\frac{r_{\rm infl}}{r_g}=\frac{c^2t^{2/3}}{G^{2/3}M_{\rm BH}^{2/3}}= \nonumber
\\
&=&7.4\times10^6\left(\frac{M_{\rm BH}}{10^{-8}M_\odot}\right)^{-2/3}\left(\frac{t}{10^{-3}\mbox{~s}}\right)^{2/3},
\label{inflrad}
\end{eqnarray}
where $r_g=2GM_{\rm BH}/c^2$ is the gravitational PBH radius.
We see that the PBH influence becomes relatively
strong at low masses $M_{\rm BH}$ and long times $t$.
The DM mass within the radius of influence is
\begin{equation}
M_{\rm DM}(t)\simeq M_{\rm BH}\left(\frac{t}{t_{\rm eq}}\right)^{1/2}=2\times10^{-8}M_{\rm BH}\left(\frac{t}{10^{-3}\mbox{~s}}\right)^{1/2},
\end{equation}
where $t_{\rm eq}\approx2.4\times10^{12}$~s is the transition time of the
Universe from the radiation-dominated cosmological
stage to the dust-like stage. In the region of influence
the particles move in the PBH gravitational field,
while outside the region of influence the cosmological
expansion continues, though, of course, this separation
is approximate and, in reality, there is a transition
region. Note that the radius of the cosmological
horizon at the radiation-dominated stage $r_H=2ct$ is
close to the radius of influence $r_{\rm infl}$ only near the PBH
formation time, while later $r_H=2ct$ expands faster
than $r_{\rm infl}$. Therefore, all of the processes we consider
occur on scales much smaller than the size of the
cosmological horizon.

The density distribution around a PBH could
be accurately calculated through numerical hydrodynamic
simulations similar to the simulations of
PBH formation \cite{p13,p32,p34,p8}. However, if the phenomena are
considered not in the immediate vicinity of the PBH
formation time but some time after, when the wave
processes will damp out, then the approximation of
quasi-stationary accretion can be used \cite{p31,p2}. Before the recombination
epoch, photons are often scattered by baryons and
are thermalized. This leads to two effects. First, a
bulk flow velocity toward the PBH appears in such
a continuous medium, though individual photons
are not captured by the PBH. Second, although the
expansion in the near zone is not the Friedmann
one, partial density equalization near the PBH and
at great distances occurs due to the existence of a
high pressure. The radiation density near the PBH is
largely determined by the density at great distances,
while the local density growth near the PBH driven
by its gravity is smoothed out strongly. Equalization
must occur at distances from the PBH smaller
than the sound horizon $r\ll r_s=2ct/\sqrt{3}$, which
is close in order of magnitude to the cosmological
horizon. At these distances, the approximation of
quasi-stationary accretion \cite{p31,p2} can be used for estimates.

\begin{figure}[t]
\begin{center}
\includegraphics[angle=0,width=0.45\textwidth]{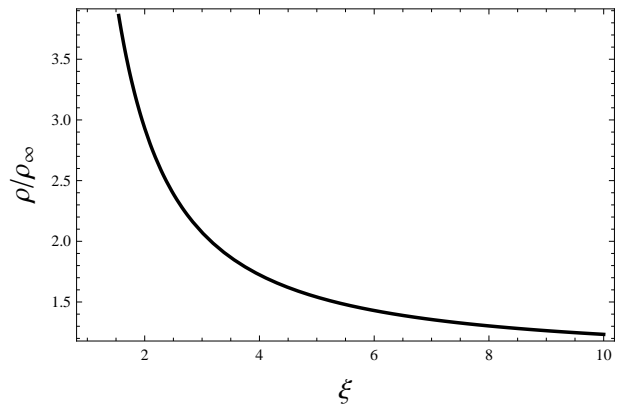}
\end{center}
\caption{ Density of a gas with the equation of state $p=\rho c^2/3$ near a black hole versus radial variable $\xi=r/r_g$ in the approximation of quasi-stationary accretion.} \label{grrho}
\end{figure}

According to \cite{p2}, the distribution
of an accreting fluid with the equation of state
$p=\rho c^2/3$ is
\begin{equation}
 \label{sol1}
 \rho=\rho_{\infty}(t)
 \left[z+\frac{1}{3(1-1/\xi)}\right]^2,
\end{equation}
where $\xi=r/r_g$,
\begin{equation}
 z=\left\{ \begin{array}{ll}
 2{\sqrt{\frac{a}{3}}}\,\cos\left(\frac{2\,\pi }{3}
 -\frac{\omega}{3}\right),& 1\leq \xi\leq3/2,\\
 2{\sqrt{\frac{a}{3}}}\,\cos\left(\frac{\omega}{3}\right),& \xi>3/2,
\end{array} \right.
\end{equation}
\begin{equation}
 \omega=\arccos\left[\frac{b}{2\,(a/3)^{3/2}}\right],
\end{equation}
\begin{equation}
 a=\frac{1}{3{\left( 1 - \frac{1}{\xi}\right) }^2},\;
 b=\frac{2}{27{\left(1-\frac{1}{\xi}\right) }^3}- \frac{27}{4\left(1-\frac{1}{\xi}\right) \xi^4}.
\end{equation}
The function (9) is shown in Fig.~\ref{grrho}.

The formalism developed by \cite{p2}
allows the velocity in the flow $u\equiv dr/ds$ to be found:
\begin{equation}
 \label{flux2}
 4u\xi^2\left(\frac{\rho}{\rho_{\infty}}\right)^{3/4}=-A,
\end{equation}
where $A=2\times3^{3/2}$. At great distances $\xi\gg 3$, the
solution (\ref{sol1}) has asymptotics $\rho\simeq\rho_{\infty}(t)/(1-1/\xi)^2$,
i.e., the density differs little from the mean cosmological
density. In this case, according to (\ref{flux2}), the
hydrodynamic flow velocity $v\sim c/\xi^2$. This quantity
is much smaller than the DM particle velocities
that we will consider below. Thus, in the Newtonian
region $r\geq10r_g$ before the kinetic decoupling
of DM particles, the density growth and the velocity
anisotropy may be neglected. In contrast, in the case
of $M_{\rm BH}>40M_\odot$, the PBH is formed already after
kinetic decoupling, and the DM density distribution
is not related to the radiation density growth around
the PBH even in the near zone.

Consider the diffusive outflow of photons from a
region of enhanced density, the Silk effect (see, e.g.,
\cite{p21}), which leads to an additional
smoothing of the radiation and DM mass
excess around the PBH before kinetic decoupling.
The photon mean free path is$l_{re}=1/(n_e\sigma_{\rm T})$, where
$\sigma_{\rm T}$ is the Thomson cross section, and the electron
number density in the cosmic plasma is
\begin{equation}
n_e\simeq\frac{\rho_{\rm eq}\Omega_b}{m_p}\frac{t_{\rm eq}^{3/2}}{t^{3/2}},
\end{equation}
$\Omega_b\approx0.045$. The Silk length $\lambda_{\rm S}\simeq(l_{re}r_H)^{1/2}$ in dimensionless
units is
\begin{equation}
\frac{\lambda_{\rm S}}{r_g}=1.5\times10^5\left(\frac{t}{10^{-3}\mbox{~s}}\right)^{3/4}\left(\frac{M_{\rm BH}}{10^{-8}M_\odot}\right)^{-1},
\end{equation}
i.e., the Silk effect can smooth out and reduce the
radiation density in the central region of a future
DM halo. This smoothing region has a size that is
smaller than the total halo size by several orders of
magnitude.


\section{Streaming of dark matter particles after kinetic decoupling}
\label{razletsec}

Let us now consider the DM density growth
around a PBH after the time of kinetic decoupling $t_d$,
when the DM particles become free. The velocity
distribution of DM particles far from the PBH is
\begin{equation}
f(\vec v)d^3v=\frac{m^{3/2}}{(2\pi
kT)^{3/2}}e^{-\frac{mv^2}{2kT}}d^3v,
\label{maxdistr}
\end{equation}
where
\begin{equation}
T(t)=T_d\frac{t_d}{t} \quad \mbox{~~at~} \quad t>t_d
\label{ttdfrac}
\end{equation}
in view of the decrease in the momentum of free particles
$p\propto 1/a(t)$. Near the PBH at distances$r\leq 10r_g$,
the distribution of particles differs noticeably from
(\ref{maxdistr}) due to the increase in radiation density compared
to the homogeneous cosmological background and
because of the existence of a bulk flow velocity toward
the PBH. However, we restrict our analysis to the
regions with$r\geq 10r_g$ in which, as was shown in Section~\ref{evolsec}, these corrections are insignificant. Therefore,
we will use (\ref{maxdistr}) in our subsequent calculations.

\begin{figure}[t]
\begin{center}
\includegraphics[angle=0,width=0.4\textwidth]{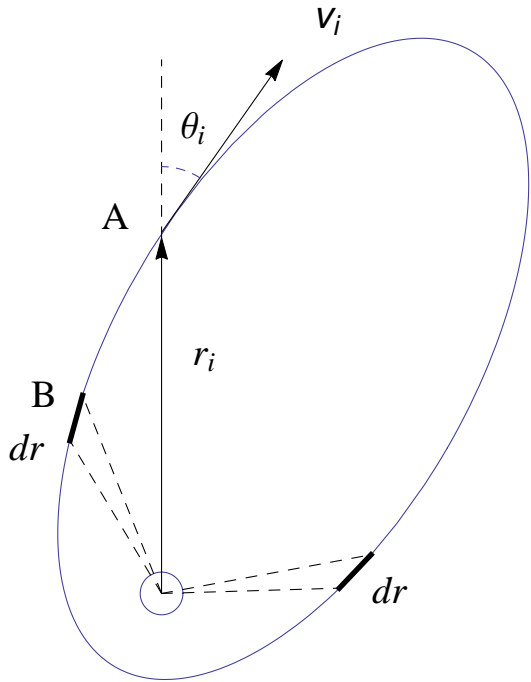}
\end{center}
\caption{An example of a particle orbit around a PBH
passing through point B. The contribution of all such
orbits to the DM density at point B at distance  $r$ from
the center is calculated. The vector $\mathbf{r_i}$ indicates the initial
position of the particle at the instant it was within the
radius of influence of the black hole, while $\mathbf{v_i}$ indicates the
particle velocity at this instant.} \label{grell}
\end{figure}

Let the PBH under consideration be at the coordinate
origin. Denote the initial distance of some DM
particle from the center by $r_i$ and its initial velocity
by $\vec v_i$. The particle energy is then $E=mv_i^2/2+U(r_i)$, where $U(r)=-GmM_{\rm BH}/r$. If the particle has
an angular momentum $l=mr_iv_i\sin\theta_i$ (see Fig.~\ref{grell}),
then the eccentricity of its orbit is \cite{p29}
\begin{equation}
e=\sqrt{1+\frac{2El^2}{G^2M_{\rm BH}^2m^3}}.
\label{eeqv}
\end{equation}
Let us consider some point B in Fig.~\ref{grell} at distance $r$
from the center and find the conditions that some
particle from the initial distribution (\ref{maxdistr}) will be in a
finite orbit around the PBH after kinetic decoupling
and will contribute to the DM density at point B. The
first condition $E<0$ means that the initial velocity is
less than the escape velocity,
\begin{equation}
v_i<\left(\frac{2GM_{\rm BH}}{r_i}\right)^{1/2}.
\label{viusl}
\end{equation}
The second condition implies that the distance $r$
lies between te minimum and maximum particle distances
from the center,
\begin{equation}
r_{\rm min}=a(1-e)\leq r \leq r_{\rm max}=a(1+e),
\label{minmaxcond}
\end{equation}
where the semimajor axis of the orbit is (Landau and
Lifshitz 1988)
\begin{equation}
a=\frac{GmM_{\rm BH}}{2|E|}.
\end{equation}
The double condition (\ref{minmaxcond}) after transformations takes
the form
\begin{equation}
\sqrt{1+\frac{2El^2}{G^2M_{\rm BH}^2m^3}}\geq\left|1+\frac{2Er}{GM_{\rm BH}m}\right|.
\label{2usl}
\end{equation}
Let us introduce the notation
\begin{equation}
x=\frac{r}{r_i}, \quad \gamma=\frac{GM_{\rm BH}}{r_iv_i^2},
\end{equation}
(\ref{2usl}) will then be written as
\begin{equation}
\cos^2\theta_i\geq\cos^2\theta_m=2x(x-1)\gamma+1-x^2.
\label{cosusl}
\end{equation}

The particle in its orbital motion traverses the segment
of radial distances from $r$ to $r+dr$ (see Fig.~\ref{grell})
twice in the orbital period
\begin{equation}
T_{\rm orb}=\frac{\pi GM_{\rm BH}m^{3/2}}{2^{1/2}|E|^{3/2}},
\end{equation}
Therefore, the particle spends the fraction $2dt/T_{\rm orb}$ of
its time at distances from $r$ to $r+dr$, where $dt$ is the
time it takes for the particle to be displaced from $r$ to
$r+dr$. Given the initial DM density $\rho_i(r_i)$, the final
density $\rho(r)$ can be written as the relation
\begin{equation}
\rho(r)4\pi r^2dr=\int 4\pi r_i^2dr_i \rho_i(r_i)\int d^3vf(v)\frac{2(dt/dr)}{T_{\rm orb}}dr,
\label{rhoitog}
\end{equation}
where the derivative $dt/dr$ is found from the equation
of motion for a particle in an orbit (Landau and Lifshitz
1988),
\begin{equation}
\frac{dt}{dr}=\frac{1}{\sqrt{2m[E-U(r)]-l^2/r^2}},
\end{equation}
while, according to the results of Section~\ref{evolsec}, we assume
the initial density $\rho_i(r_i)$ at distances $r\geq 10r_g$
to be approximately uniform and equal to the cosmological
DMdensity:
\begin{equation}
\rho_i(r_i)\simeq\rho_d\frac{t_d^{3/2}}{t^{3/2}},
\label{rhoieq7}
\end{equation}
where
\begin{equation}
\rho_d\simeq\Omega_m\rho_{\rm eq}\left(\frac{t_{\rm eq}}{t_d}\right)^{3/2}\simeq4.7\times10^3\mbox{~g~cm$^{-3}$}
\end{equation}
and $\Omega_m\simeq0.27$. The collisionless system under consideration
is described by the Liouville equation, and
the method being applied in this paper is equivalent
to an approximate solution of this equation. Indeed,
Eq.~(\ref{rhoitog}) expresses the density conservation law in
phase space integrated over the momenta by taking
into account the volume transformation in momentum
space, which follows from the Liouville equation.

When integrating in (\ref{rhoitog}) over $r_i$, we should separately
consider the regions with $r_i\leq r_{\rm infl}(t_d)$ and
$r_i>r_{\rm infl}(t_d)$, where the radius of influence is given by
Eq.~(\ref{inflrad1}). The first region at the time $t_d$ is entirely in
the region of PBH influence, and the particles in this
region have a common velocity distribution (\ref{maxdistr}) with
$T=T_d$ and a common density $\rho_i(r_i)\simeq\rho_d$. In contrast,
at $r_i>r_{\rm infl}(t_d)$, the region of influence gradually
expands. For each radius $r_i$, the temperature $T$ and
density $\rho_i$ are found from Eqs.~(\ref{ttdfrac}) and (\ref{rhoieq7}), respectively,
in which $t$ is specified by the equation $r_{\rm infl}(t)=r_i$. It should be noted that the kinetic decoupling
of DM particles occurs not instantaneously, and the
scatterings of particles during the transition period
can slightly change the final density of the captured
DM particles.

Inequalities (\ref{viusl}) and (\ref{cosusl}) separate out the region
in parameter space over which the integration in (\ref{rhoitog})
is performed. It is convenient to divide this integral
into two parts with $x<1$ and $x\geq1$. The internal
integration over the velocity directions, i.e., over the
angles $\cos\theta_i$, is done analytically, while the remaining
double integrals over the initial radii $r_i$ and the
absolute values of the initial velocities $v_i$ are found
by numerical methods. The maximum possible radii
$r_i$ are assumed to be equal to the radius of influence
given by (\ref{inflrad}) at the time $t=t_{\rm eq}$. At $t=t_{\rm eq}$, a DM
mass equal to the PBH mass is inside the region
of PBH influence, and a DM halo is subsequently
formed around the PBH by the mechanism of secondary
accretion, when the PBH no longer determines
the entire gravitational field but serves only as
a small perturbation.

\begin{figure}[t]
\begin{center}
\includegraphics[angle=0,width=0.45\textwidth]{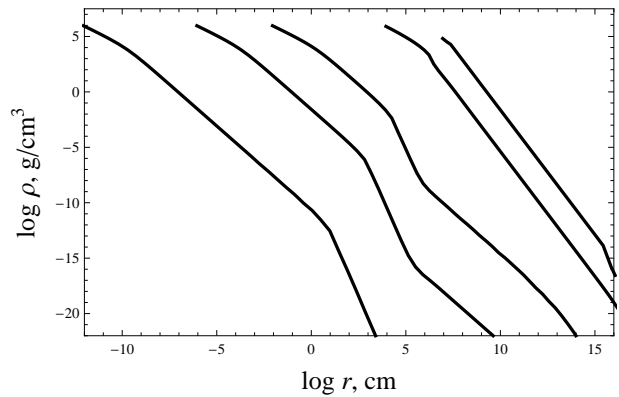}
\end{center}
\caption{DM density around a PBH versus radius $r$ for the following PBH masses (from left to right):
$M_{\rm BH}=10^{-18}$, $10^{-12}$, $10^{-8}$, $10^{-2}$, and $10M_\odot$.} \label{grdeniteq}
\end{figure}

The results of our numerical calculations for various
PBH masses are shown in Fig.~\ref{grdeniteq}. The numerical
algorithm constructed in this paper gives an acceptable
accuracy only in the range of masses $M_{\rm BH}\sim(10^{-18}-1)M_\odot$. The resulting density at small distances
from the PBH exceeds $\rho_d$. This means that
particles with low angular momenta are in eccentric
orbits approaching the PBH at small radii. A universal
behavior of the density at small radii, where
there are segments with a density profile close to the
power-law one $r^{-1}$, is also seen in Fig.~\ref{grdeniteq} at $M_{\rm BH}\leq10^{-2}M_\odot$, but at large $r$ the profile experiences a
break due to the change of the regime of DM halo
formation. The radii in Fig.~\ref{grdeniteq} are shown formally
starting from $r=3r_g$. Strictly speaking, our calculations
performed within the framework of Newtonian
dynamics are applicable only at $r\geq10r_g$. Therefore,
the density at smaller radii must be considered as an
estimate. In the next section, we will show that the
density in the central region of the halo plays no role,
because the DM in the central spikes has strongly
annihilated by now and the density has decreased by
several orders of magnitude.


\section{Early particle annihilation in spikes}
\label{earlyannsec}

If the DM particles are able to annihilate, then
their density will decrease with time. As was shown
in \cite{p5} and  \cite{p39}, the maximum DM density in a particular
object at the present time does not exceed
\begin{eqnarray}
&& \rho_{\rm max}\simeq \frac{m}{\langle\sigma_{\rm ann} v\rangle t_0} \simeq 9.4\times10^{-15}\left(\frac{m}{70\mbox{~GeV}}\right)\times
\label{annwr}
\\
&\times &\left(\frac{\langle\sigma_{\rm ann} v\rangle}{3\times10^{-26}\mbox{~cm$^3$s$^{-1}$}}\right)^{-1}
\!\!\!\left(\frac{t_0}{1.4\times10^{10}\mbox{~years}}\right)^{-1}\mbox{~g~cm$^{-3}$},
\nonumber
\end{eqnarray}
where $t_0$ is the time elapsed since the formation of
the object. The thermal production cross section
for DM particles in the early Universe $\langle\sigma_{\rm ann} v\rangle\simeq3\times10^{-26}$~cm$^3$~s$^{-1}$ is taken as a normalization for the
annihilation cross section (see, e.g.,  \cite{p20}).

It follows from Fig.~\ref{grdeniteq} that the density of the DM
halo for all of the PBH masses considered at small
radii exceeds considerably (\ref{annwr}). This means that the
dense central regions of the halo existed only at early
epochs, while by now the DM density in the halos
around PBHs has decreased to (\ref{annwr}). An accurate calculation
of the law of decrease in central density due
to particle annihilation is a more complicated problem
that is beyond the scope of this paper. Particles
with different orbital parameters annihilate at each
point of the density spike. Therefore, a self-consistent
allowance for the decrease in DM density simultaneously
in all regions of the density spike is needed
for the calculation. Such a calculation is planned to
be performed in future works. At early times, the
annihilation-generated gamma-ray emission experienced
absorption and thermalization in the cosmic
plasma. The annihilation that continued during primordial
nucleosynthesis or at the reionization epoch
of the Universe could influence these processes, but
the quantitative role of this influence requires a separate
study. DM annihilation around PBHs is an
additional factor that can lead to chemical anomalies
in the baryonic halos around PBHs considered by
 \cite{p19}.

Thus, DM density spikes, from which halos with
a central density $\rho\sim \rho_{\rm max}\sim10^{-14}$~g~cm$^{-3}$ and a
decreasing density on the periphery have been left at
present, existed around PBHs. The sizes of these
halos are equal in order of magnitude to the radii of
PBH influence given by Eq.~(\ref{inflrad1}) at $t=t_{\rm eq}$.


\section{Annihilation of dark matter around PBHs at the present epoch, observational constraints}

Consider the annihilation of DM particles in
density spikes around PBHs that are located in our
Galaxy at the present epoch. The annihilation in
a spike around a single PBH, i.e., the number of
annihilated particles per unit time
\begin{equation}
\dot N=4\pi\int r^2dr\rho^2(r)\frac{\langle\sigma_{\rm ann}v\rangle}{m^2},
\label{sepfour}
\end{equation}
where the profiles obtained in Section~\ref{razletsec} by taking
into account the early annihilation in the central part
considered in Section~\ref{earlyannsec} are used as the density profile
in the spike $\rho(r)$. Thus, starting from a radius of $\sim3r_g$,
we assume that $\rho(r)\sim\rho_{\rm max}=10^{-14}$~g~cm$^{-3}$, while
at large radii, when the densities in Fig.~\ref{grdeniteq} decrease to
$\rho_{\rm max}$, the profiles shown in Fig.~\ref{grdeniteq} are used under the
integral in (\ref{sepfour}).

After the beginning of the dust-like stage of the Universe
at $t> t_{\rm eq}$, a DM halo begins to grow around the
PBH at distances $r>r_{\rm infl}(t_{\rm eq})$ through the mechanism
of secondary accretion  \cite{p7}. Its
density distribution is
\begin{equation}
\rho(r)\simeq3\times10^{-21}\left(\frac{r}{1\mbox{~pc}}\right)^{-9/4}\left(\frac{M_{\rm BH}}{10^2M_\odot}\right)^{3/4}\mbox{g~cm$^{-3}$},
\label{rhoihsa}
\end{equation}
with the outer boundary of (\ref{rhoihsa}) being determined
by the influence of ordinary inflationary density perturbations,
so that the total mass of the DM halo
around the PBH exceeds the PBH mass $M_{\rm BH}$ approximately
by two orders of magnitude  \cite{p15}. The density (\ref{rhoihsa}) does not
exceed the halo density at $r=r_{\rm infl}(t_{\rm eq})$. Therefore, the
outer halo (\ref{rhoihsa}) makes a minor contribution to (\ref{sepfour}),
while the central region of the halo with density (\ref{annwr})
and the parts of the halo adjacent to it shown in Fig.~\ref{grdeniteq}
make a major contribution.

The total annihilation signal from some direction
characterized by the angle $\psi$ with respect to the
Galactic center is
\begin{equation}
J_\gamma=2\eta_{\pi^0}\dot N \frac{\Omega_{\rm BH}}{\Omega_m M_{\rm BH}}\int dL\rho_{\rm H}(r(L)),
\label{jgam}
\end{equation}
where $\Omega_{\rm BH}$ is the cosmological density parameter for
PBHs with masses $M_{\rm BH}$, $\eta_{\pi^0}\sim10$ is the number of
photons per $\pi^0$ decay, and the integration is along
the line of sight. The hadronic annihilation channel,
when most of the gamma-ray photons are emitted
during the decays of neutral pions $\pi^0\to2\gamma$ produced
by the annihilation of DM particles, is assumed to be
the main one. As the density profile of the Galactic
halo $\rho_H(r)$, we use the Navarro--Frenk--White profile
 \cite{p33}
\begin{equation}
 \rho_{\rm H}(r)=
 \frac{\rho_{0}}{\left(r/R_s\right)\left(1+r/R_s\right)^2}.
 \label{halonfw}
\end{equation}
where $R_s=20$~kpc and $\rho_{0}=6.7\times10^6M_\odot$~kpc$^{-3}$.

\begin{figure}[t]
\begin{center}
\includegraphics[angle=0,width=0.45\textwidth]{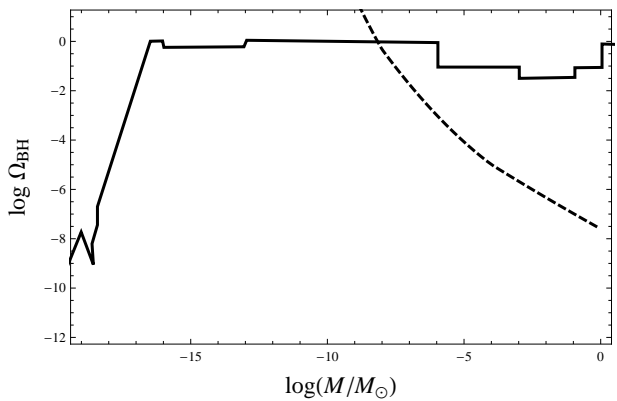}
\end{center}
\caption{The solid curve indicates the known upper bounds on the cosmological PBH density parameter $\Omega_{\rm BH}$ from Carr
et al. (2010). The dashed curve indicates the constraints based on the DM particle annihilation effect obtained here.
} \label{grom}
\end{figure}

\begin{figure}[t]
\begin{center}
\includegraphics[angle=0,width=0.45\textwidth]{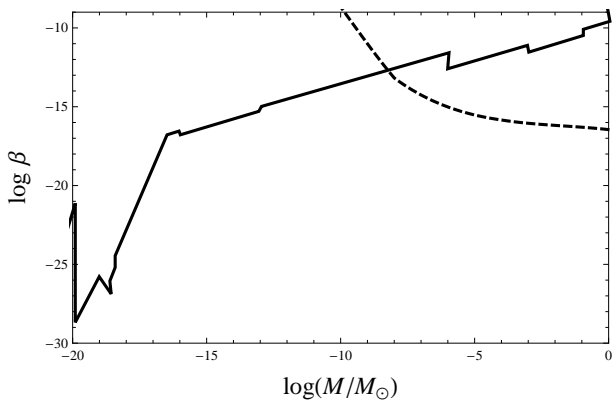}
\end{center}
\caption{Same as Fig.~\ref{grom} but for the fraction $\beta$ of the mass of the Universe gone into PBHs at the time of their formation.
} \label{grbet}
\end{figure}

Let us compare (\ref{jgam}) calculated toward the Galactic
anticenter $\psi=\pi$ (this gives the minimum signal
and, accordingly, the most conservative constraint)
with the Fermi-LAT observational constraint
from the diffuse gamma-ray background
$J^{\rm obs}(E>m_{\pi^0}/2)=1.8\times10^{-5}$~cm$^{-2}$~s$^{-1}$~sr$^{-1}$  \cite{p1}. The condition $J_\gamma<J^{\rm obs}$ gives an
upper bound on $\Omega_{\rm BH}$, which is shown in Fig.~\ref{grom}
together with several other known constraints on
PBHs from  \cite{p14}. In particular, these
are the constraints from the Hawking radiation and
microlensing and the constraint from the overall
cosmological PBH density. These constraints are
often expressed via the fraction $\beta$ of the mass of the
early Universe gone into PBHs at the time of their
formation (\ref{th}). The quantity $\beta$ is related to $\Omega_{\rm BH}$ as
 \cite{p14}
\begin{equation}
\Omega_{\rm BH}\simeq5\times10^{17}\beta\left(\frac{M_{\rm BH}}{10^{15}\mbox{~g}}\right)^{-1/2}.
\label{ombeteq}
\end{equation}
The constraint on $\beta$ following from neutralino annihilation
in density spikes is shown in Fig.~\ref{grbet}.

We see from Figs.~\ref{grom} and ~\ref{grbet} that the constraint from
annihilation at $M_{\rm BH}\geq 10^{-8}M_\odot$ gives constraints on
the number of PBHs that are several orders of magnitude
more stringent than other known constraints.


\section{Conclusions}

DM clumps can be produced by various mechanisms
 \cite{p6}. They can be
formed both from cosmological density perturbations
in the dark matter itself and around compact
seed masses, for example, around cosmic strings
 \cite{p27} or PBHs  \cite{p15,p35,p30,p36,p28,p38}. Secondary accretion, the
infall and virialization of cold DM onto PBHs, was
thought to be the main mechanism for the formation
of DM clumps around PBHs. However, we showed in
this paper that there exists another mechanism that
gives rise to denser DM clumps around PBHs than
was considered in secondary accretion models.

The DM density around PBHs grows at the
radiation-dominated stage due to the presence of
slow DM particles in their thermal velocity distribution.
Fairly slow particles after their kinetic
decoupling are in finite orbits around PBHs and
produce high-density DM clumps. Considering the
kinematics of particles around PBHs allowed the
density profile to be found. The DM particles in the
central regions of clumps have managed to annihilate
by now. However, the remaining halos are still very
dense, and intense annihilation occurs in them. This
effect can be interesting for experiments on indirect
detection of DM particles though the search for their
annihilation products, because the annihilation of
particles in spikes can contribute to the observed
gamma-ray emission. Comparison of the calculated
signal with the Fermi-LAT observational limits gives
upper bounds on the present-day cosmological PBH
density parameter from $\Omega_{\rm BH}\leq1$ to $\Omega_{\rm BH}\leq10^{-8}$,
depending on the PBH masses at $M_{\rm BH}\geq10^{-8}M_\odot$.
Comparable (in magnitude) but weaker constraints
$\Omega_{\rm BH}\leq 10^{-4}$ were obtained previously in  \cite{p28}, where the density profile in a spike
was assumed to be $\rho\propto r^{-3/2}$.

However, it should be noted that our constraints
are largely model-dependent ones: they depend
fundamentally on the as yet unknown properties
of DM particles. The derived constraints refer to
standard neutralinos or to other DM particles having
the properties of weakly interacting massive particles
(WIMPs), i.e., having masses and annihilation
cross sections comparable to them in order of magnitude.
For other DM particles, both the velocity
distribution (\ref{maxdistr}) and the annihilation signals can
be significantly different. For example, if the DM
particles do not annihilate at all, then the formation
of a DM density spike around PBHs is still possible,
but, in this case, there is no early annihilation and
no decrease in central density and the signals are
absent in the cosmic gamma-ray emission. The
density spikes can be bright gamma-ray sources only
at certain masses and annihilation cross sections of
DM particles. The derived constraints do not refer,
for example, to the models in which the DM 
consists of PBHs. Nevertheless, the neutralinos in
nonminimal supersymmetric models so far remain
among the most probable DM candidates, and the
constraints obtained here can hold.

Author is grateful to V.K.~Dubrovich for useful
discussions.


\end{document}